# High Performance Fin-FET electrochemical sensor with high-k dielectric materials

Serena Rollo[a,b], Dipti Rani[a], Wouter Olthuis[b], César Pascual García[a]

[a]Nano-Enabled Medicine and Cosmetics group, Materials Research and Technology Department, Luxembourg Institute of Science and Technology (LIST), Belvaux, Luxembourg
[b]BIOS Lab on Chip Group, MESA+ Institute for Nanotechnology, University of Twente, Enschede, The Netherland

**ABSTRACT**

In this work we combine a Fin Field Effect Transistor (Fin-FET) characterised by a high height to width aspect ratio with high-k dielectric materials to study the optimized design for chemical-FETs to provide higher transconductance (and thus a better signal to noise ratio), increased dynamic range and chemical stability. We used pH sensing to verify the design. We explored the sensitivity and response linearity of silicon dioxide, alumina and hafnium oxide as dielectric materials sensing pH, and compared their chemical stability in different acids. The high aspect ratio fin geometry of the sensor provides high currents, as well as a planar conduction channel more reliable than traditional silicon nanowires. The hafnium oxide Fin-FET configuration performed the best delivering the most linear response both for the output and transfer characteristics providing a wider dynamic range. Hafnium oxide also showed the best chemical stability. Thus, we believe that the developed high aspect ratio Fin-FETs/high-k dielectric system can offer the best compromise of performance of FET-based sensors.

1. **INTRODUCTION**

Bio-Field Effect Transistors (Bio-FETs) are FET based sensors combined with a biological recognition element able to sense biomolecules. They are an interesting alternative for label free detection of biomarkers in the fields of genomics[1-3] and proteomics[4-6] for applications in medical diagnostics, drug discovery and basic research, offering multiplexing capability, portability and miniaturisation, real-time analysis, selectivity, low cost. Despite these desirable features, there is not yet a portable, low cost device in the market based on this technology. In fact, there are challenges to overcome when scaling up from the laboratory to the industry level related to the reliability of the performance among devices, the functionalization with the bio-recognition element and the chemical stability of the surface[7-9] in particular for applications that require an extended contact of the sensor surface with the sample fluid. To improve the performance of bio-FETs and chemical-FETs in general the original design of planar devices evolved to nano-sensors like nanowires[10], and new materials were introduced to provide an increase in the transduced signal and chemical stability of the interface[11-13]. Owing to the miniaturisation achieved by nanowires, the sensitivity of label free sensing increased from μM to fM and the incubation time needed for heavy molecules to reach the equilibrium decreased from days to hours or minutes[14,15]. Nevertheless the improved sensitivity of nano devices came at the cost of impacting negatively the signal to noise ratio and the variability of the current signal among devices[16,17]. Recently we proposed a Fin-FET design with a high aspect ratio of the height to width (>10), in which the width of the sensor was comparable to that of nanowires, but due to the bigger height, it resulted in a planar conduction channel[18]. This change in the geometry improved the signal to noise ratio, the linearity of the output signal and provided a higher surface area, which is favourable for the reliability of the functionalisation as compared to nanowires. The device design provides a compromise to increase the total signal, while providing a good response time for assays at low concentrations, for which the sensing is diffusion limited[19].

The dielectric interface of the FET in contact with the electrolyte is a key component of the sensor as it determines its chemical stability[20] as well as the transduction. It can be used as receptor for simple molecules or ions in solution such as protons[21,22] or as the support for the functionalisation of biorecognition layers that improve the selectivity of the sensor[23]. In the case where the interface is directly used to capture molecules, the surface chemical properties of the dielectric itself determine the surface potential that regulates the conductivity of the transistor across the source to drain channel. The conduction is also affected by the dielectric constant ($k$) of the material that determines the capacitance effect between the sensor surface and the conduction channel. The fabrication of devices with silicon dioxide ($SiO_2$) as dielectric interface is convenient but it is not preferable since $SiO_2$ has low pH buffer capacity in comparison to other dielectric materials, suffers from drift, hysteresis, leakage currents, penetration of ions when in contact with the electrolyte for an extended period of time[24,25]. Other dielectrics such as aluminium oxide ($Al_2O_3$)[11,12,26] and hafnium oxide ($HfO_2$)[12,27] can be used to improve the sensor properties, being more resistant to ion penetration and providing a higher dielectric constant that increases the transconductance further by increasing the capacitive effect in the semiconductor even with physically thicker layers. Combining the design of a high aspect ratio Fin-FETs with high-k dielectrics can enhance their specific advantages, improving the superior linear response of the output current and increasing the sensitivity and signal to noise ratio by improving the transconductance responsible of the signal transduction. Materials with better chemical performance meaning higher intrinsic buffer capacity, while also being more resistant to dissolution in both acidic and basic conditions, have the potential to provide reliability and stability to the device.

To measure the impact of the dielectric in FETs, the detection of the acidity of a solution in aqueous electrolytes (pH) has been used as a direct comparison of the performance among different oxides[28,29]. The response of the dielectric towards pH can be described using the combined Gouy-Chapman-Stern and Site-Binding (GCS-SB) models, where the GCS model describes the electrical double layer that forms at the oxide interface, and the SB model describes the grade of ionization (protonation or deprotonation) of the surface chemical groups of the dielectric barrier[30]. Using both models it is possible to derive the relationship between the bulk pH and the potential at the oxide surface ($\Psi_0$), characterised by the oxide sensitivity $\Delta\Psi_0/\Delta pH$, which determines the chemical response of the material. Silicon oxide shows pH sensitivities of 20 to 40 mV/pH depending on the quality of the grown layer, and a nonlinear response in a wider pH range due to its low intrinsic buffer capacity[31-34]. *$Al_2O_3$* and *$HfO_2$* have shown sensitivities equal or higher than 55 mV/pH, and improved linearity in a wide pH range[11,12,28,35]. An ultimate design of a FET sensor has to combine the sensor geometry with the impact of the dielectric material in the transduction and surface properties in the chemical performance (sensitivity and stability).

In this work we combine a p-doped high aspect ratio Fin-FET design with different dielectrics as thermally grown *$SiO_2$* and atomic layer deposited *$Al_2O_3$* and *$HfO_2$* on a thin *$SiO_2$* adhesion layer. We have studied the pH sensitivity in terms of variations of $\Psi_0$, which we relate to the intrinsic properties of the material (dissociation constants of the surface active groups and surface density of the surface reactive sites). We also compare the effects of transducing the variations of $\Psi_0$ within two similar Fin-FET devices with *$SiO_2$* and *$HfO_2$*, respectively. Using a Nernst-Poisson model[18], we calculate the effective dielectric constant of the stack *$SiO_2/HfO_2$*. Finally, we test the stability of the three oxides comparing a controlled citric acid buffer with natural citrus juices. We proved that while *$Al_2O_3$* represents an improvement to *SiO₂*, *HfO₂* provides the best chemical stability in time and overall enhances the transduction properties of the Fin-FETs. Owing to the combination of the high aspect ratio of the sensors configuration with the high-k and chemically stable *HfO₂* we report the highest performance of this electrochemical sensor.

## 2. MATERIALS AND METHODS

**2.1 Silicon Fin-FETs fabrication**

We fabricated silicon Fin-FETs by anisotropic wet etching of p-doped silicon on insulator (SOI) substrates with a 2 ± 0.1 and 3 ± 0.1 μm thick silicon device layer (<110> oriented) with resistivity of 0.115 Ω·cm (equivalent doping $10^{17}$ /cm$^3$) and a 1 μm thick buried *SiO₂* procured from Ultrasil Corporation. The substrates were diced in chips of 1 x 1 cm$^2$ before starting the fabrication of the Fin-FETs. Briefly, we used Maskless photolithograpy (MLA 150 Heidelberg Instruments) and e-beam lithography (FEI Helios electron microscope) on the negative resist ma-N 2403 to pattern lines with widths ranging from 400 to 700 nm on a thermally grown thin *SiO₂*, oriented along the <110> direction parallel to the primary flat of the substrate in order to get the desired shape after wet etching. The Fin-FET shape originates from the different rates at which the <110> and the <111> planes are etched. The device lateral walls lay on the <111> planes. The etching along the vertical direction (<110> plane) is about 10 times faster than along the <111> planes. Knowing the plane dependent etching rates[36] and device layer thickness of Si, lithography mask was designed with defined line widths to have final wire width on the chips. The connection between the lines and the contact pads was achieved through approaching pads with a triangular footprint designed at the angles of ≈54.7° and 35.3° with respect to the primary flat to provide a smooth profile between the channel and the pads after etching. This pattern was then transferred by Reactive Ion etching through a CF₄ process of 15 minutes at a pressure of 75 mTorr and power of 25 W to the undelaying oxide. The samples were then treated with HF to remove the excess of oxide outside the lithographed area and to obtain a smooth surface. The anisotropic etching was achieved with a 25 % wt Tetramethylammonium hydroxide, 8.5 %vol of Isopropanol water solution lasting for ≈23 and ≈30 minutes for the complete etching of the 2 and 3

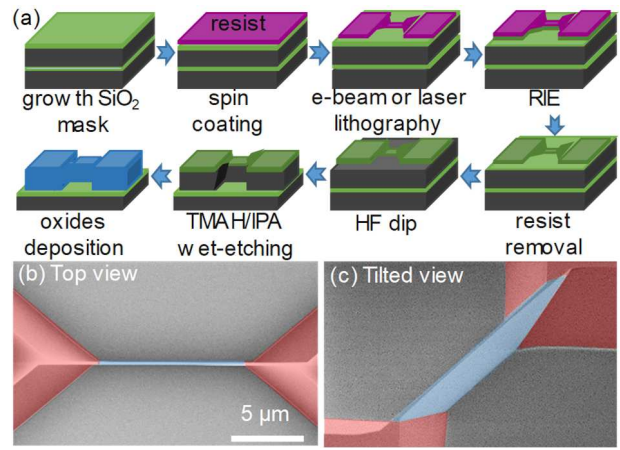

**Figure 1** (a) Schematic representation of the fabrication process of Fin-FETs on SOI substrates based on laser or e-beam lithography on a negative resist, and wet etching in a TMAH/IPA wet etching solution. Silicon is represented in dark grey while silicon oxide is represented in green. The resist is depicted in purple and the final oxide as sensing layer in blue. (b) and (c) Top and tilted SEM pictures respectively of one representative Fin-FET device after fabrication. The silicon body of the device standing on the grey buried oxide and the contacts are shadowed in blue and red, respectively.

μm thick substrates, respectively. After a 1 minute dip in HF to remove the $SiO_2$ mask, samples were ready for the deposition of the gate oxide stacks. We used 20 nm of thermally grown $SiO_2$, and 10 nm atomic layer deposited (ALD) $Al_2O_3$ and $HfO_2$ with 7 nm of $SiO_2$ thermally grown as interlayer between the silicon and the ALD grown oxides to have the pH sensitive layers. Figure 1 (a) schematically shows the fabrication steps. Figure 1 (b) and (c) show Scanning Electron Microscope (SEM) pictures of a representative device with top and tilted views respectively. The high aspect ratio Fin-FET channel is shadowed in blue between the source and drain contact pads which are shadowed in red. The ohmic contacts and the leads necessary for the integration into a plastic circuit board (PCB) were defined by optical lithography on regions of the devices shadowed in red part. The ohmic contacts were a Ti/Al/Au stack (2/160/5 nm) e-beam evaporated, while the leads were Au 150 nm. Another lithography step on an epoxy (SU8) allowed to open windows on the Fin-FET region while protecting the contacts. After wire bonding to the PCBs dipstick, the samples were protected with a medical grade epoxy glue (Loctite EA M-31CL, Henkel). The final devices had a length of 14 μm at the middle of the Fin-FET and width ranging from 150 to 400 nm. Each chip contained eleven Fin-FETs. Table 1 summarizes the characteristics of thickness of the deposited oxide ($t_{ox}$), average width (w) of the devices on the same chip with the same oxide, and height (h) of the fabricated devices.

**Table 1** Characteristics of thickness of the deposited oxide ($t_{ox}$), width (w) and height (h) of the fabricated devices.

| Device oxide | $t_{ox}$ (nm) | w (nm) | h (nm) |
|---|---|---|---|
| $SiO_2$ | 20 | ≈170 | 2 |
| $Al_2O_3$ | 10 | ≈400 | 2 |
| $HfO_2$ | 10 | ≈200 | 3 |

## 2.2 pH sensitivity characterization

Experiments of pH sensitivity were carried out in buffer solutions with pH from 3 to 11 in step of 1. The buffers were prepared by mixing a solution of *$KH_2PO_4$*, citric and boric acids at 0.1 M all, with a *$KNO_3$* 0.1 M solution in equal volume proportion, for a final pH of 2.5. More basic pH buffer solutions were obtained by addition of a 0.1 M solution of *KOH*. All the solutions were prepared using Milli-Q water as solvent. With this procedure, the total ionic strength remained constant at 0.1 M. For the electrochemical characterization the chips were immersed into the buffer solutions with a calomel reference electrode (BioLogic R-XR300) for biasing the electrolyte and a commercial pH meter (Sentron SI600) to check the pH throughout the measurements. We used a Keithley 2614HB DC source meter to apply the voltage between the source and drain contacts and to the reference electrode and a multiplexer Keithley 3706A System Switch/Multimeter connected to a switching box to characterize the devices in sequence.

## 2.3 Measurements of acidity in citrus juices

First, we prepared a solution of citric acid 0.01M by dissolving 0.48 g in 250 mL of Milli-Q water. The resulting solution had an acidic pH of 3 measured with a commercial pH meter. The lemon and orange juices were obtained from freshly squeezed fruits and filtering the pulp. Their pH was also measured with the pH meter. For the measurements of citric acid we used the same set-up described in the paragraph above.

## 3. RESULTS AND DISCUSSION

### 3.1 Surface sensitivity of Fin-FETs with $SiO_2$, $Al_2O_3$, $HfO_2$

In order to determine the surface sensitivity of the grown oxides, we measured the transfer characteristics, source drain current ($I_{ds}$) *vs.* reference electrode voltage ($V_{ref}$), at constant source drain

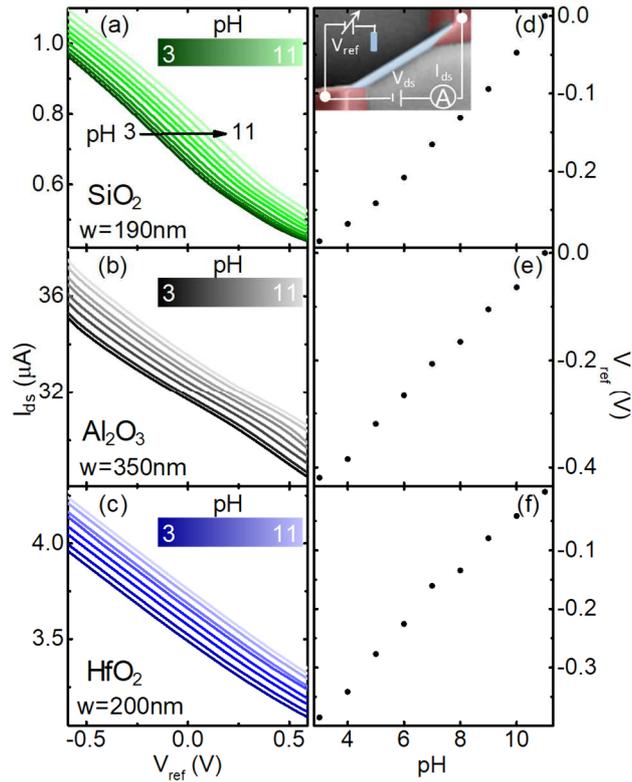

**Figure 2** (a) to (c) Examples of transfer characteristics $I_{ds}$ vs $V_{ref}$ at fixed $V_{ds}$ for three representative devices with the three different oxides. The curves were measured in buffers at pH 3 to 11 represented with a coloured scales. The width of the devices is referred as w. The inset in (d) represent a schematic of the measurement setup. (d) to (f) $V_{ref}$ vs pH measured from the curves in (a) to (c) from which the sensitivity of the oxide was evaluated as shifts of $V_{ref}$ at each pH to maintain a constant current.

voltage ($V_{ds}$). The pH sensitivity ($\Delta V_{ref}/\Delta pH$) was evaluated from the shifting of the transfer curves at a constant current with different buffers pH values. The variations of the reference electrode voltage ($\Delta V_{ref}$) compensate (and correspond to) the changes in the surface potential ($\Delta \Psi_0$) induced by the different proton concentrations. The choice of $V_{ds}$ followed from a preliminary characterization performed at neutral pH and $V_{ref}$ = 0 V. For this characterization we measured the output current $I_{ds}$ vs. $V_{ds}$ from all devices. At higher $V_{ds}$ values we observed the pinching off of the carrier density in the channel by the deviation of the current characteristics from the linear behaviour. We restricted the study to the linear range of $I_{ds}$ versus $V_{ds}$ in order to be able to explain the variation of the conductance of the device with the ohmic contribution of the conducting channel cross section, and its dimensions. To this objective, for the characterization we used a $V_{ds}$ of 0.1 V for the narrower devices (like the ones reported for *SiO₂* and *HfO₂*), and 0.5 V for the wider ones (like the ones reported for *Al₂O₃*), while $V_{ref}$ was swept in a range between -0.6 and 0.6 V in all cases. The transfer characteristics were acquired in a pH range between 3 and 11 in steps of 1 by immersing the samples into the buffer solutions. Multiple Fin-FETs on three different chips having the three oxides as pH sensitive layers were characterized with the same procedure. The inset of fig. 2 (d) reports a scheme of the measuring set up. Figure 2 (a) to (c) show the transfer characteristics of three representative devices from each type family of dielectrics at different values of pH represented on coloured scales as shown in fig. 2 (a) to (c). The width of the tested devices (w) is also specified. In each case we observed a shifting of the transfer characteristics toward more positive $V_{ref}$ while moving from acidic to basic buffers. This is because the majority carriers in the semiconductor channel are holes affected by $\Psi_0$. When the pH increases there are less protons interacting with the oxide surface, thus lower $\Psi_0$ compared to more acidic conditions, and therefore higher $V_{ref}$ is required to compensate the electrostatic potential at the oxide liquid interface to maintain a constant current flowing through the channel. Figures 2 (d) to (f) show the shift of $V_{ref}$ with pH in the curves in fig. 2 (a) to (c) derived as the $V_{ref}$ necessary to keep the value of $I_{ds}$ at $V_{ref}$ = 0 V constant from the value at pH 11, which corresponds to the variations of $\Psi_0$ due to the different proton concentrations. The relation between the surface potential and the pH is derived by combining the electrostatic interactions at the dielectric surface and the distribution of ions inside the electrolyte starting from the oxide surface, which was found earlier[30]:

$$\frac{\Delta \Psi_0}{\Delta pH_B} = -2.303 \frac{kT}{q} \alpha \qquad \text{Eq. 1}$$

Where $pH_B$, $k$, $T$ and $q$ represent the pH in the bulk electrolyte, the Boltzmann constant, the absolute temperature and the elementary charge, respectively. α is a sensitivity parameters with a value varying between 0 and 1 depending on intrinsic properties of the oxide. For α=1 the sensor has a so called Nernstian sensitivity of 59.2 mV/pH at 298 K. We obtained an estimation of the sensitivity of the different oxides from the linear fit of the curves like the ones showed in fig. 2 (d) to (f) acquired from all the devices, obtaining the average value of the sensitivity, and the standard deviation for each type of oxide. We found that the response of the dielectrics to different proton concentrations, which experimentally translates into a shift of the transfer characteristics at different pH values, were qualitatively similar among Fin-FETs with the same oxide. *Al₂O₃* provided the best performance in terms of sensitivity with 54.2 ± 1.9 mV/pH, while the one for *HfO₂* was 49.8 ± 0.6 mV/pH. For both oxides the experimental results are in agreement with other values of sensitivities reported in literature[11,12,26,27]. While *Al₂O₃* and *HfO₂* have an approximately linear response in the pH range considered, *SiO₂* has a lower sensitivity in acidic conditions compared to basics due to the lower intrinsic buffer capacity of the oxide surface at low pH where the groups at the surface interacting with the protons in electrolyte are close to saturation and are not able to buffer the changes of proton concentration. In the pH range between 6 and 11, where silicon oxide has the highest sensitivity, we estimated a value of 42.1 ± 0.5 mV/pH. Close to saturation (i.e. at the point of zero charge of the oxide surface, pH$_{pzc}$) at pH lower than 6, we estimated a sensitivity of 30.2 ± 1.1 mV/pH. These values are also in agreement with other values reported in literature[31-34].

The origin of the different pH sensitivities among the different oxides can be explained in terms of the acidic and basic dissociation constants ($K_a$ and $K_b$, respectively) of the active groups from each oxide surface, and the surface density of surface reactive sites ($N_s$). The combination of the Site Binding model which describes the reactivity of the surface groups with the Gouy-Chapman-Stern model, which describes the formation of an electrical double layer at the oxide/electrolyte interface gives an expression for the sensitivity parameter α in eq. 1 containing the differential capacitance $C_{diff}$ and the intrinsic buffer capacity $β_{int}$.

$$\alpha = \frac{1}{\frac{2.303kTC_{diff}}{q^2 \beta_{int}}+1} \qquad \text{Eq. 2}$$

The differential capacitance depends on the electrolyte (solvent dielectric constant and ionic strength), while $β_{int}$ depends on $K_a$, $K_b$, and $N_s$ and is linked to the ability of the oxide to buffer small changes of surface charge[30]. Higher values of $β_{int}$ are related to more reactive surfaces, thus improved sensitivities. From the experimental pH sensitivities, we evaluated α using Eq. 1. We also estimated $C_{diff}$ using the estimation of $C_{diff}$ presented in literature by Van Hal et al. that modeled $C_{diff}$ as the series capacitance of the Stern capacitance $C_{St}$ (the contribution of the layer of charges in closest contact with the oxide) and the diffuse layer capacitance $C_{DL}$ (from Gouy and Chapman)[30]. $C_{St}$ has been theoretically calculated for different ionic strengths of the electrolyte[37], and we used the same value considered by Van Hal et al. of 0.8 F/m². For the estimation of $C_{DL}$ it is assumed that the total charge in the diffuse layer ($\sigma_{DL}$) is equal to the charge at the oxide surface ($\sigma_0$), which yields the expression for $C_{DL}$ derived by Van Hal et al.[30]:

$$\sigma_{DL} = -(8kT\varepsilon_0\varepsilon_w n^0)^{1/2} sinh\left(\frac{zq\Psi_0}{2kT}\right) = -C_{DL}\Psi_0 = -\sigma_0 \qquad \text{Eq. 3}$$

Where $\varepsilon_0$, $\varepsilon_w$ and $n^0$ are the vacuum and water relative permittivities and the number concentration of each ion of the electrolyte, respectively. Using eq. 3 we calculated the experimental $C_{diff}$ for an electrolyte with a 0.1 M ionic strength as the one we used in our experiments, and combining with the experimental sensitivity in eq. 2 we calculated the experimental buffer capacity $\beta_{exp}$. We compared $\beta_{exp}$ with the intrinsic buffer capacity calculated using literature values of the acidic and basic dissociation constants of the surface reactive groups ($K_a$, $K_b$), and surface density of surface reactive sites ($N_s$)[28,30,38] according to the expression for $\beta_{int}$ given by Van Hal et al., as reported in S.I. We obtained $\beta_{exp}$ of 0.6 x 10¹⁸, 1.5 x 10¹⁸ and 1.7 x 10¹⁸ groups/m², for $SiO_2$, $Al_2O_3$ and $HfO_2$ respectively, compared to the theoretical values of 0.9 x 10¹⁸, 3.7 x 10¹⁸ and 2.8 x 10¹⁸ groups/m², respectively. In each case we noticed that the experimental values of intrinsic buffer capacity are lower than the theoretical ones. The difference may be attributed to the different way the oxides are grown, the presence of impurities on the oxide surface and defects coming from the deposition step that affect the total number of reactive sites.

## 3.2 Relevance of Fin-FETs integration with high-k dielectrics

Higher k dielectrics yield improvements to the sensor. Fin-FETs with high aspect ratio show more linear and higher transconductance, $\Delta I_{ds}/\Delta pH$ respect to SiNWs[18]. We expect that higher k dielectrics will further improve the output characteristics of these devices. We compared the conductance of two Fin-FETs devices with the two oxides having the most different dielectric constants, meaning the ones with 20 nm thermally grown silicon oxide and the one with approximately the same total thickness, but with half of the material made by hafnium oxide. Both devices had approximately the same base width and length (190 nm and 14 μm, respectively), and heights of 2.16±0.1 and 2.90±0.1 μm for the *SiO$_2$* and *HfO$_2$*, respectively (measured by profilometry).

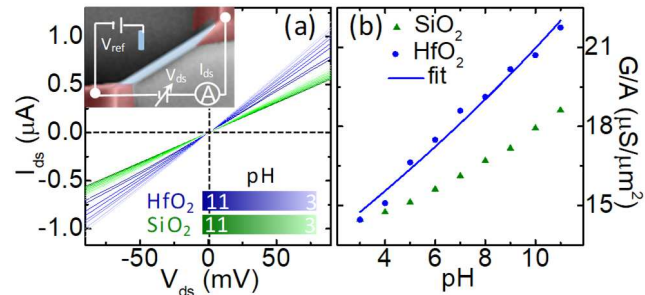

**Figure 3** (a) Output characteristics $I_{ds}$ vs $V_{ds}$ at fixed $V_{ref}$ = 0 V for the *SiO$_2$* and *HfO$_2$* Fin-FETs respectively. The curves at different pH are coloured according to the scales in the inset. A schematic of the measuring setup is also shown as inset. (b) Normalized conductance vs pH for the Fin-FETs having approximately the same width with *SiO$_2$* and *HfO$_2$* as pH sensitive layers, calculated from the curves in fig. 3 (a). The fitting of the experimental data with the Nernst-Poisson model used to estimate the dielectric constant of the deposited *HfO$_2$* is represented as a blue line.

The output characteristics $I_{ds}$ vs $V_{ds}$ were explored in a pH range from 3 to 11 with $V_{ref}$ = 0 V. We choose this value of $V_{ref}$ since at neutral pH $I_{ds}$ was linear in a range of $V_{ref}$ between -200 and 200 mV, which is the variation of surface potential expected in the considered pH range. Thus $I_{ds}$ can be described with the ohmic contribution of the non-depleted region with a Nernst-Poisson model. $I_{ds}$ was acquired sweeping $V_{ds}$ between -100 and 100 mV. $I_{ds}$ had a linear behaviour in that range as shown in fig. 3 (a) for the devices with *SiO$_2$* and *HfO$_2$* using colour scales for pH between 3 and 11 with steps 1. The measuring setup is also shown schematically in the inset of fig. 3 (a). From the data in fig. 3 (a) we estimated the conductance (G) as the slope of the linear fittings. In fig. 3 (b) we report the conductance of the two Fin-FETs with the different sensing oxides normalized by the cross section of the device, to take into account the difference in height between both devices and allow a comparison. The experimental data are represented as green triangles and blue dots for the devices with *SiO$_2$* and *HfO$_2$*, respectively. The conductance increases towards more basic pH values in both cases as $I_{ds}$ depends on the surface potential $\Psi_0$ which depends on the proton concentration as already explained. The variation of the conductance we obtained as 522 ± 12mS/pH and of 912 ± 19mS/pH per unit area for the $SiO_2$ and the $HfO_2$ devices respectively. The effect of enhanced variation of the conductance in the device with $HfO_2$ is due to the contribution from the higher intrinsic sensitivity of the material $\Delta\Psi_0/\Delta pH$, and to the higher dielectric constant, which increases the transconductance in the device. The higher linear response of the $HfO_2$ pH surface sensitivity is transferred to the output response. The higher sensitivity offered by $HfO_2$ through the whole acidity range combined with the high aspect ratio fin geometry of the sensor channel offers better performances in a wider dynamic range.

We estimated the dielectric constant of the $HfO_2$ layer ($\varepsilon_{HfO2}$) using a Nernst-Poisson model to fit the experimental data (blue line in fig. 3 (b)) combined with the experimental sensitivity parameter $\alpha$ retrieved from the $\Delta V_{ref}/\Delta pH$. The model is based on eq. 1 to describe dependence of $\Psi_0$ with pH that modulates the Poisson distribution of charges determining the depleted region that lastly controls the output current (a more detailed description can be found in our previous article[18]). The effective dielectric constant $\varepsilon_{eff}$ of the $SiO_2/HfO_2$ stack was modelled with two capacitors in series from each oxide layer with known thicknesses ($t_{SiO2}$= 7 nm and $t_{HfO2}$ =10 nm measured during the growth in a dummy sample with ellipsometry):

$$\varepsilon_{eff} = \frac{(t_{SiO2}+t_{HfO2})\varepsilon_{SiO2}\varepsilon_{HfO2}}{t_{SiO2}\varepsilon_{HfO2}+t_{HfO2}\varepsilon_{SiO2}} \quad \text{Eq. 4}$$

Considering the dielectric constant for $SiO_2$ $\varepsilon_{eff}$ = 3.9 we obtained $\varepsilon_{eff}$ of 7.4 and thus from eq. 4 $\varepsilon_{HfO2} \approx$ 20, which is in agreement with other values in literature for ALD deposited $HfO_2$[35,39]. The integration of high aspect ratio Fin-FETs with high-k materials provided the best performance in the output currents of the sensors for linearity and sensitivity.

### 3.3 Stability of the oxides in different acidic media

To study the stability over time of the different oxides in contact with fluids, we used the Fin-FETs with the three different interfaces to sense the acidity of squeezed lemon and orange juices where the main component responsible of the acidity is citric acid (7%, and 4-5%, concentration for lemon and orange juice, respectively). We compared the behaviour of the devices in citric juices with a 0.01 M citric acid buffer (pH 3) monitoring the fluctuations of the output current while moving the sensors

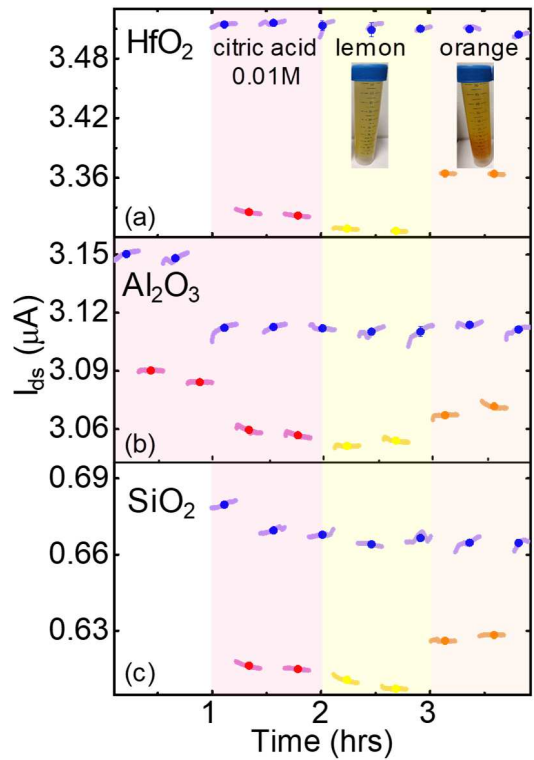

**Figure 4** (a) to (c) Output current $I_{ds}$ vs time from three Fin-FETs devices with the three different oxides in different media represented in different colours. The current was recorded for 15 minutes after each change of fluid after five minutes of stabilization time. In the same graphs the average value of $I_{ds}$ in the 15 minutes of the measurements is represented as coloured dots. Pictures of the citrus juices used in the experiments are shown as an inset.

from one liquid to another. We tested a family of devices for each oxide with a common external reference electrode in each run moving the devices alternatively between water and the other acid solutions every 15 minutes and waiting five minutes before starting the next measurement to allow the stabilization of the sensor. To avoid cross contamination the sensors were rinsed with deionized water and blow dried with nitrogen in between each exchange of solutions. Figure 4 shows the output currents and average values (using dots) for each cycle of the same devices shown in fig. 2. The water, citric acid buffer, lemon and orange juices solutions are represented with blue, red, yellow and orange colours, respectively. The insets shows pictures of the juices liquid samples used in the experiments. *HfO₂* showed a very reproducible behaviour throughout the measurements. The current returned to approximately the same values depending on the pH of the solution with a drift < 10 nA (~5% of the measured range) along three hours of measuring time. This was not the case for *Al₂O₃* and *SiO₂*. Specifically the current in the device with *Al₂O₃* showed an abrupt change of about 40 nA ( > 60%) after

1 hour followed by an stabilisation. Then the device behaved similarly to the SiO$_2$ one, which in three hours had a drift of 12 nA (~15%). We attribute the abrupt change in the *Al$_2$O$_3$* to the corrosion by citric acid which provokes the detachment of material especially in the pH range 3-6 as reported in literature[40]. After the *Al$_2$O$_3$* layer was totally corroded the *SiO$_2$* beneath was exposed stabilizing the device. In the transfer characteristics recorded after the experiment we noticed a decrease of the pH sensitivity in line with values reported for *SiO$_2$* (transfer characteristics reported in SI), which support our hypothesis. In the *SiO$_2$* device the drift during the first hour is attributed to the intrinsic drifting normally observed in silicon oxide[24,41]. *SiO$_2$* suffers issues of ions reactions and incorporation when in contact with electrolyte for an extended period of time which affects the oxide stability until an equilibrium is reached between the reactive groups at the oxide surface and ions in the solution, and the stability restored[24]. As discussed before, we obtained higher total average sensitivity for the *HfO$_2$* Fin-FETs with a ΔR/R = 6.9% (R refers to resistance of the device) between pH 7 and 2.8 compared to that of the *SiO$_2$* FinFETs of ΔR/R = 5.9% in the same range (after normalization to the cross section to take into account for the different heights).

4. **CONCLUSIONS**

In this work we investigated the surface sensitivity of different dielectric materials and the way they influence the transconductance in high aspect ratio Fin-FET chemical sensors. The chemical affinity of the different hydroxyl groups at the surface of the dielectrics provides the surface sensitivity of the material, which was tested by acidity measurements in a pH range from 3 to 11. We obtained surface sensitivities of 54.2 ± 1.9mV/pH, 49.8 ± 0.6 and 37.5 ± 1.3mV/pH for *Al$_2$O$_3$*, *HfO$_2$* and *SiO$_2$* respectively. While *Al$_2$O$_3$* and *HfO$_2$* had an approximately linear variation of the surface potential throughout the range investigated (pH 3-11), *SiO$_2$* showed a lower sensitivity in acidic conditions attributed to the saturation of the reactive groups on the surface at low pH, next to the pH$_{PZC}$. We evaluated the experimental intrinsic buffer capacity (β$_{int}$) of the three oxides observing the poorer sensitivity of *SiO$_2$* among the three oxides. We also investigated the effect of *SiO$_2$* and *HfO$_2$* on the transconductance of the Fin-FETs and observed an almost doubled response for the *HfO$_2$*, which we attribute to the enhanced surface sensitivity of the material as well as to the higher dielectric constant. This high aspect ratio Fin-FET/*HfO$_2$* dielectric combination allows to increase the linearity of the output current with the concentration of the analyte and thus the dynamic range of the devices.

We investigated the stability of the three oxides when exposed to liquids for a long period of time monitoring the fluctuations of the output currents of the three Fin-FET families of oxides. We measured the acidity of different liquids other than ideal buffer solutions, such as citrus juices, where the acidity is mainly provided by the citric acid. In the device with *HfO$_2$* the output current was stable, coming back at the same value after each change of the media. Along the three hours of the experiments we measured a drift of less than 5% of the measured range. For the device covered with *Al$_2$O$_3$* we observed an abrupt change of more than 60% of the measured range after one hour, which we attributed to the corrosion of the material by the citric acid. The device with *SiO$_2$* showed a drift of 15% of the measured range in the first hour, attributed to reactions of ions at the surface and ion incorporation, while the stability was restored after one hour.

In conclusion combining the Fin-FET geometry which intrinsically benefits an improved linearity in the transduction due to the 2D depletion along the width of the device, with high-k materials providing higher transconductance and chemical stability, improves the FET/dielectric material system offering higher performances of sensitivity and linearity of the response to provide wider dynamic ranges, and long term stability in liquid environment. These properties are all desirable features for biosensing applications and FET based biosensors development.

**CONFLICTS OF INTEREST**

There are no conflicts to declare.

**ACKNOWLEDGEMENTS**

We would like to thank Dr. Sivashankar Krishnamoorthy for useful discussions and help during the project.

**FUNDING**

This project was financed by the FNR under the Attract program, fellowship number 5718158 NANOpH.